\documentclass[referee,pdflatex,sn-mathphys]{sn-jnl}

\jyear{2022}%

\begin{document}

\title[Solving TOV Equations in $f(T)$ Gravity]{Solving Tolman-Oppenheimer-Volkoff Equations in $f(T)$ Gravity: a Novel Approach Applied to Polytropic Equations of State}


\author*[1]{\fnm{José Carlos N.} \sur{de Araujo}}\email{jcarlos.dearaujo@inpe.br}
\equalcont{These authors contributed equally to this work.}

\author[1]{\fnm{Hemily G.M.} \sur{Fortes}}\email{hemily.gomes@gmail.com}
\equalcont{These authors contributed equally to this work.}

\affil*[1]{\orgdiv{Divisão de Astrofísica}, \orgname{Instituto Nacional de Pesquisas Espaciais}, \orgaddress{\street{Avenida dos Astronautas 1758}, \city{São José dos Campos}, \postcode{12227-010}, \state{São Paulo}, \country{Brasil}}}


\abstract{The Teleparallel Theory is an alternative theory of gravity equivalent to General Relativity (GR) and with non-vanishing torsion $T$. Some extensions of this theory, the so-called $f(T)$ models, have been subject of many recent works. The purpose of our work in the end is to consider recent results for a specific family of $f(T)$ models by using their corresponding Tolman-Oppenheimer-Volkof equation to describe compact objects such as neutron stars. By performing numerical calculations, it is possible to find, among other things, the maximum mass allowed by the model for a neutron star for a given equation of state (EOS), which would also allow us to evaluate which models are in accordance with observations. {To begin with,} the present work, the second in the series, considers polytropic EOSs since they can offer a  simpler and satisfactory description for the compact objects. In addition, with these EOSs, we can already assess how different the $f(T)$ theories are in relation to GR with respect to the stellar structure. {Moreover, there is only a few studies in $f(T)$ regarding polytropic EOSs, therefore there is space for further studies.}
The results already known to GR must be reproduced to some extent and, eventually, we can find models that allow higher maximum masses than GR itself, which could explain, for example, the secondary component of the event GW190814. This particular issue is subject of another paper, the third in the series, where realistic EOSs are considered.}

\keywords{Alternative theories of gravity, Neutron star}

\maketitle

\section{Introduction}
\label{int}

The Teleparallel equivalent of General Relativity (TEGR)\cite{Moller}-\cite{Unzi} is an alternative description of gravity first introduced by Einstein himself. This theory is defined on a Witzenböck space–time, which is curvatureless, but with a non–vanishing torsion. Just as modifications that can be made in General Relativity (GR) leading us to the $f(R)$ theories, it is possible to extend the telleparallel to the so-called $f(T)$ gravity where one uses a function of torsion as the Lagrangian density. These alternative theories of gravity have, in general, simpler field equations than those obtained in $f(R)$ and have presented interesting cosmological and astrophysical solutions \cite {Ferraro}-\cite{Karami}.

Despite some theoretical issues \cite{PCC,brax}, GR has been a great success as a theory of gravity. Therefore, it is expected that a well conceived alternative model of gravity retrieve the results of GR in some limit and also offer some advantage or new physical possibilities. In this sense, it is always relevant to reproduce for the alternative models some analysis already done in the context of GR. Thus, our main interest here is to study the behaviour of spherically symmetric stars in some $f(T)$ models. More specifically, in \cite{F-A1} we have obtained the correspondent Tolman-Oppenheimer-Volkoff (TOV) equations \cite{TOV} for a set of  ${f(T)}$ {models.} TOV equations are derived from the theory of gravity and can be used to model the structure of a spherically symmetrical object in gravitational equilibrium, for instance, a neutron star. Solving these equations for a given equation of state, it is possible, for example, to determine the maximum mass allowed for the object. 

Considering different gravity models could, in principle, lead us to different maximum masses {for a given equation of state} (see, e.g., \cite{Nunes}). This would be important, for example, to explain one of the components involved in the event GW190814 \cite{GW} concerning the coalescence between a black hole of 22.2–24.3 $M_\odot$ and a compact object of mass 2.50–2.67 $M_\odot$, the latter being a relatively high-mass neutron star or a low-mass black hole.

{The TOV equations in \cite{F-A1} for the ${f(T)}$ {models} were found without any restriction on the metric functions or perturbative calculations}, leading to a nontrivial coupled differential equations. The next step is to perform the numerical calculations by choosing a specific equation of state and the suitable boundary conditions. For the purposes of this work, the equation of state to be considered will be the polytropic equation \cite{eos0} since it {can} offer a simpler and satisfactory description for neutron stars, for example. The studies related to realistic EOSs are left for another paper to appear elsewhere.

{Notice that although the $f(T)$ literature is quite vast, there are not many papers that adequately deal with the modelling of spherically symmetric stars, even regarding polytropic EOSs. }

{The appropriate approach for modelling of polytropics in $f(T)$ is relatively recent in the literature.} {We refer the reader to \cite{Ilijic18} for further details. These authors, however, considered for a given $f(T)$
only a particular polytropic EOS. Moreover, their calculations stop converging for some combinations of parameters, which was interpreted as an indication that there are no solutions for these combinations.} {In our previous paper \cite{F-A1}, we present a novel approach, in which we obtain a system of equations different from \cite{Ilijic18}. In this novel approach, we confirm the lack of solutions for some combinations of parameters studied by \cite{Ilijic18}. However, we obtained solutions for certain combinations of parameters that \cite{Ilijic18} failed to find due to instabilities in their numerical calculations.} {Therefore, our approach is more robust,} {as it opens up the possibility of investigating a domain of parameters that Ref. \cite{Ilijic18} is unable to investigate. This is, in itself, a relevant novelty of our approach.}

{Since \cite{F-A1} and \cite{Ilijic18} consider together only two distinct polytropic indices, there is room in the literature to explore other ones.}

{Notice that polytropic EOSs are also interesting in the study of stellar stability in $f(T)$, which is still an open problem. This is therefore an additional motivation to study them.} {Thus, we present in Section \ref{sec 3}, three additional polytropic EOSs, not studied yet in the literature.}

A brief summary of the results obtained in \cite{F-A1}, in particular the set of differential equations to be solved for the {${f(T) = T + \xi T^2}$ models}, will be presented in Section \ref{sec 2}. All this analysis, the methodology used and the results obtained are presented in Section \ref{sec 3}. In Section \ref{sec 4}, we present the final remarks.

\section{Equations of Stellar Structure in f(T) Models}\label{sec 2}

 In theories with torsion, {such as $f(T)$,}  the dynamical variable is no longer the metric $g_{\mu\nu}$, but rather the tetrad ${e^a}_\mu$ which has to be carefully chosen for each specific physical context \cite{TT,review2,Tamanini}. {In the covariant formulation of $f(T)$ gravity \cite{Krssak}, besides the tetrad the spin connection must be also considered as a dynamical variable. Consequently, in this covariant formulation one does not need to be worried about the tetrad, since anyone is equally good.{\footnote{{Following Ref \cite{Tamanini}, good tetrads are those that do not impose any restrictions in the form of $f(T)$. If any restriction is imposed in the form of $f(T)$, you have a bad tetrad.}}}}  Choosing a given tetrad, one only has to obtain from it its corresponding spin connection. {It is well also known that the spin connection corresponding to tetrad adopted in our approach is null. We refer the reader to Refs. \cite{F-A1} and \cite{Krssak} for more details. Thus, our analysis is tetrad independent since it is consistent with the covariant formulation.}
 
{Since here we are interested in modelling stars without rotation, therefore spherically symmetrical, the appropriate spacetime can be given by the metric} 
{$ds^2=e^{A(r)}\, dt^2-e^{B(r)}\, dr^2-r^2\, d\theta^2-r^2 \sin ^2 \theta \, d\phi^2$.}

The equations of motion can be obtained from the action, 
\begin{eqnarray}
{{ S = \int   \ \left( \frac{f(T)}{16\pi} + \mathcal{L}_m \right) \ {\rm det}({h_\mu}^a)\ d^4 x \ ,}}
\end{eqnarray}
where the form of $f(T)$ {is the same considered in our previous paper} \cite{F-A1}, namely,
\begin{eqnarray}
 f(T)=T+\xi \, T^2 \ ,\label{fT}
\end{eqnarray}
where $\xi$ is an arbitrary real. Notice that, for $\xi=0$, it is expected to retrieve the TEGR results, which will be a reference parameter for the numerical calculations in the next section. {Additionally, $\mathcal{L}_m$ is the Lagrangian density of matter fields. Notice that the variation regarding matter fields leads to the usual energy-momentum tensor. In this paper, the source term is given by an isotropic perfect fluid, with pressure $P$ and energy density $\rho$.}

Then, one obtains nontrivial coupled differential equations which mix all the functions and their derivatives, namely as $B$, $B'$, $A'$ and $A''$. In general, we find in the literature this type of equations being solved by considering some approximations, restriction of the functions or perturbative calculations \cite{Ganiou}-\cite{Pace} in order to avoid a cumbersome process. Conversely, in \cite{F-A1} we have the TOV equations for the model (\ref{fT}) written explicitly without the need of any previous assumptions.

It is worth mentioning that \cite{Ilijic18} follows a similar step, but they do not show explicitly the set of differential equations that is effectively numerically solved. {An important difference is that they do not use the {\it conservation equation}, which greatly simplifies the numerical calculations.  It seems that they have not realised that the appropriate combination of the equations presented in Section 2 of their paper leads to the {\it conservation equation}.} {Furthermore, as already mentioned, our approach is numerically more robust than theirs.}

We present below the main equations obtained in our previous paper \cite{F-A1} which will be used in the further numerical calculations. 

First, we have the expression for the pressure $P(r)$ in terms of the functions $B(r)$ and $A'(r)$:
\begin{eqnarray}
P=\frac{{\rm e}^{-B}}{8\pi r^4}\Big\{{r}^{2} \left(1-{{\rm e}^{B}} \right) + r^3A'
+12r({{\rm e}^{-1/2\,B}}-1)^2\xi A'+
\nonumber \\
+2{{\rm e}^{B}}({{\rm e}^{-1/2\,B}}-1)^3(3{{\rm e}^{-1/2\,B}}+1)\xi +
\nonumber \\
+ 2r^2 \xi({{\rm e}^{-1/2\,B}}-1)(3{{\rm e}^{-1/2\,B}}-1) \, {A'}^2\Big\} \ .\hspace{0.3cm}
\label{press}
\end{eqnarray}

Also, we have an equation for $B'(r)$ in terms of $B(r)$ itself, $A'(r)$, $\xi$ and $\rho$ (mass density):
{
\begin{eqnarray} 
B'&=& -\frac{{{\rm e}^{B}}}{r}\bigg\{{r}^{4}\big(1-{{\rm e}^{-B}}-8\,\pi \,{r}^{2}\rho\big)
-96\pi\,{r}^{4}\rho ({{\rm e}^{-1/2\,B}}-1)^2\xi+ 
\nonumber \\    
&-& 6\,{r}^{2}({{\rm e}^{-1/2\,B}}-1)^3(5+3{{\rm e}^{-1/2B}})\xi
-8({{\rm e}^{-1/2\,B}}-1)^5(11+9{{\rm e}^{-1/2B}})\xi^2 +
\nonumber \\    
&-& 8\,{r}^{3}{{\rm e}^{-1/2\,B}}({{\rm e}^{-1/2\,B}}-1)\Bigl[8\,\pi \,{r}^{2}\rho
+({{\rm e}^{-1/2\,B}}-1)(2{{\rm e}^{-1/2\,B}}+1)\Bigr]\,A'\xi+
\nonumber \\    
&-& 16\, r\, {{\rm e}^{-1/2\,B}}({{\rm e}^{-1/2\,B}}-1)^4(9{{\rm e}^{-1/2\,B}}+5)\,A'\xi^2
+2\,r^4 {{\rm e}^{-B}}({{\rm e}^{-1/2\,B}}-1)^2  \,{A'}^2\xi+
\nonumber \\    
 &-& 8\,r^2 {{\rm e}^{-B}}({{\rm e}^{-1/2\,B}}-1)^3 (9{{\rm e}^{-1/2\,B}}-1)\,{A'}^2\, \xi^2\bigg\}
\bigg/ 
 \bigg\{ r^4+16\,{r}^{2}({{\rm e}^{-1/2\,B}}-1)^2\xi+
\nonumber \\   
&+& 48\,({{\rm e}^{-1/2\,B}}-1)^4\xi^2
+16\,{r}^{3}{{\rm e}^{-1/2\,B}}({{\rm e}^{-1/2\,B}}-1) A'\xi +
\nonumber \\   
&+& 96\,r\,{{\rm e}^{-1/2\,B}}({{\rm e}^{-1/2\,B}}-1)^3 A'\xi^2 
   +48\,{r}^{2}{{\rm e}^{-B}}({{\rm e}^{-1/2\,B}}-1)^2
 {A'}^2\xi^2 \bigg\} 
\label{BL}
\end{eqnarray}
}

Additionally, in order to solve these differential equations, it will be necessary to consider the well-known {\it conservation equation} \cite{Bohmer}:
\begin{equation}
    A' = -2 \frac{P'}{P+\rho} \ .
\label{ce1}    
\end{equation}
{It is worth stressing that in \cite{Bohmer} it is shown in detail that the {\it conservation equation} is valid in $f(T)$ gravity.}

{Notice that the use of the {\it conservation equation} combined with equation (\ref{press}) leads to an algebraic equation involving $A'$, $B$ and $P$. Consequently, one does not need a differential equation for $A$, since what appears in the system of equations to be solved is $A'$, which can be obtained algebraically in our approach.}

{In the next section, some advantages of our approach compared to that of \cite{Ilijic18} become quite clear.}

Notice that mass density $\rho=\rho(r)$ appears in the equations above and it will be related to the pressure $P$ through the equation of state. Furthermore, the mass $m(r)$ can be obtained from: 
\begin{eqnarray}
\frac{dm}{dr}=4\pi \rho  r^2\ .
\label{dmdr}
\end{eqnarray}

It is worth noticing that the above equation is the same of TOV GR, which is also adopted {implicitly or explicitly} by other authors (see, e.g., \cite{Ilijic20}). Nonetheless, recall that the calculation of mass in $f(T)$ gravity is considered by some authors an open problem (see, e.g., \cite{Ilijic18} and \cite{olmo}). {Taking this issue into account, we also calculate the mass that would be measured by an ``observer at infinity"  as an alternative way to obtain the mass of a star in $f(T)$ gravity. We refer the reader to \cite{F-A1} for details.}

{The total rest mass $M_0$, although is not an observable, is an interesting quantity to calculate, since it can be used to compare different theories. For example, different theories must provide different values of $M_0$ for given equation of state and central density.

The total rest mass $M_0$ of a star is obtained by solving
\begin{eqnarray}
\frac{dm_0}{dr}=4\pi \rho_0 \,e^{B/2} r^2 \, ,
\label{dm0dr}
\end{eqnarray}
where $\rho_0$ is the rest mass density.
}

In the next section we present the numerical calculations and the corresponding modellings for the particular case of polytropic EOSs.

 \section{Numerical Calculations and Discussions}\label{sec 3}
In what follows we describe in detail how the numerical calculations have been performed. First of all, as already mentioned, in the present paper we consider polytropic EOSs, {which are widely adopted in studies of stellar structure whether in Newtonian gravitation, General Relativity or in any alternative theories of gravity in general.} Recall that the polytropic reads
\begin{equation}
    P = k\, {\rho_0}^ \gamma,
    \label{tpol}
\end{equation}
where $\rho_0$ is the rest-mass density, $k$ is the polytropic gas constant and $n$ defined by $\gamma \equiv 1+1/n$ is the polytropic index. From the first law of thermodynamics, one obtains the mass-energy density, $\rho$, namely $\rho = \rho_0 + n\,P$ (see, e.g., \cite{BS2010}).

It is also considered in the literature polytropic EOSs written as follows 
\begin{equation}
    P = k\, \rho ^ \gamma,
    \label{ppol}
\end{equation}
which is sometimes referred to as pseudo-polytropic. This is particularly useful for describing realistic EOSs in terms of polytropics, i.e., the EOS is written in piecewise polytropic form.

{It is noteworthy that even considering the same polytropic indices, the EOSs above provide completely different stellar models.}

Following \cite{BS2010}, and also \cite{F-A1}, we set all quantities in nondimensional form. In practice, this is equivalent to set
$k = G = c = 1$. This is an useful procedure since our aim here is to compare how different $f(T)$ is as compared to GR regarding compact objects.

To construct a compact star in GR, for example, one has to integrate numerically the differential equations for $m(r)$ and $P(r)$ for a given EOS. Then, it is necessary to choose a central density $\rho_c$ and, using the EOS, one obtains the central pressure $P_c$. Additionally, providing the central boundary conditions, 
\begin{equation}
    m = 0  \quad {\rm and}  \quad P = P_c \quad {\rm at} \quad r =0,
\end{equation}
one obtains $m(r)$, $P(r)$ and $\rho(r)$, i.e., the structure of the star. The radius of the star, $R$, is given by the value of $r$ for which $P(r) = 0$. That is, one starts the integration of the set of differential equations at $r=0$ and continues it untill the value of $r$ for which $P(r)=0$. {Finally, the mass of the star, calculated just like in GR, is then given by ${M_{GR} = m(R)}$.}

Here, to model a compact star, we have an additional equation, namely, the differential equation for $B(r)$. Then, one needs to provide a central boundary condition for $B$.  Based on GR, it is possible to convince oneself that $B_c ~=~0$.

{In addition, we calculate the mass as ``mass measured by an observer at infinity'' ($M_{I\! N\! F}$). In \cite{F-A1} we show in detail that this mass is given by}
\begin{equation}
    M_{I\! N\! F} = \lim_{r\to\infty} \frac{1}{2}\, r\,B(r).
    \label{minf}
\end{equation}
{That is, we also need to integrate the system of equations for the vacuum. In practice, the above equation converges for values of $r$ on the order of hundreds of the star's radius.}

Before proceeding, it is worth mentioning that we have written a numerical code in Python in order to perform the numerical integration of the set of differential equations. 

A very useful way to compare different theories of gravity by using models of compact stars is via sequences of models. That is, for a range of values of $\rho_c$, one obtains the corresponding masses and radii. Consequently, one obtains the ``Mass $\times$ Radius'' and ``Mass $\times$ $\rho_c$'' curves, from which one can obtain, for example, the maximum mass allowed for a given EOS.

Notice that $\xi$ in $f(T) = T + \xi T^2$, the particular theory we are considering, is dimensionless too, since $G = c = 1$.

{There are different motivations for choosing this particular $f(T)$. It is simple and is inspired by Starobinsky's $f(R)$ gravity model, which has the same functional form.}

The first model we consider is {essentially GR}, because it will be the reference for comparisons. However, instead of using TOV GR to generate sequences of models for given polytropic EOSs, we can set $\xi=0$ in our equations. This gives $f(T)=T$, which is nothing but the 
TEGR, whose TOV equations after a simple manipulation are reduced to TOV GR, as it should be. In fact, this is an important test to show that our scheme is consistent.

For the first set of models, we consider a pseudo polytropic EOS with index $n = 1$, which gives $P= \rho^2$, where we set $k=1$ since we are dealing with dimensionless quantities, as already mentioned. 
{It is worth stressing again that $P= \rho^\gamma$ and $P= \rho_0^\gamma$ lead to different star models. Therefore, it is also worth studying the modelling adopting pseudo polytropic EOSs. All these modellings can be helpful in future studies regarding stability of stars in $f(T)$ gravity.}

\begin{figure*}
    \centering
    \includegraphics[scale=0.24]{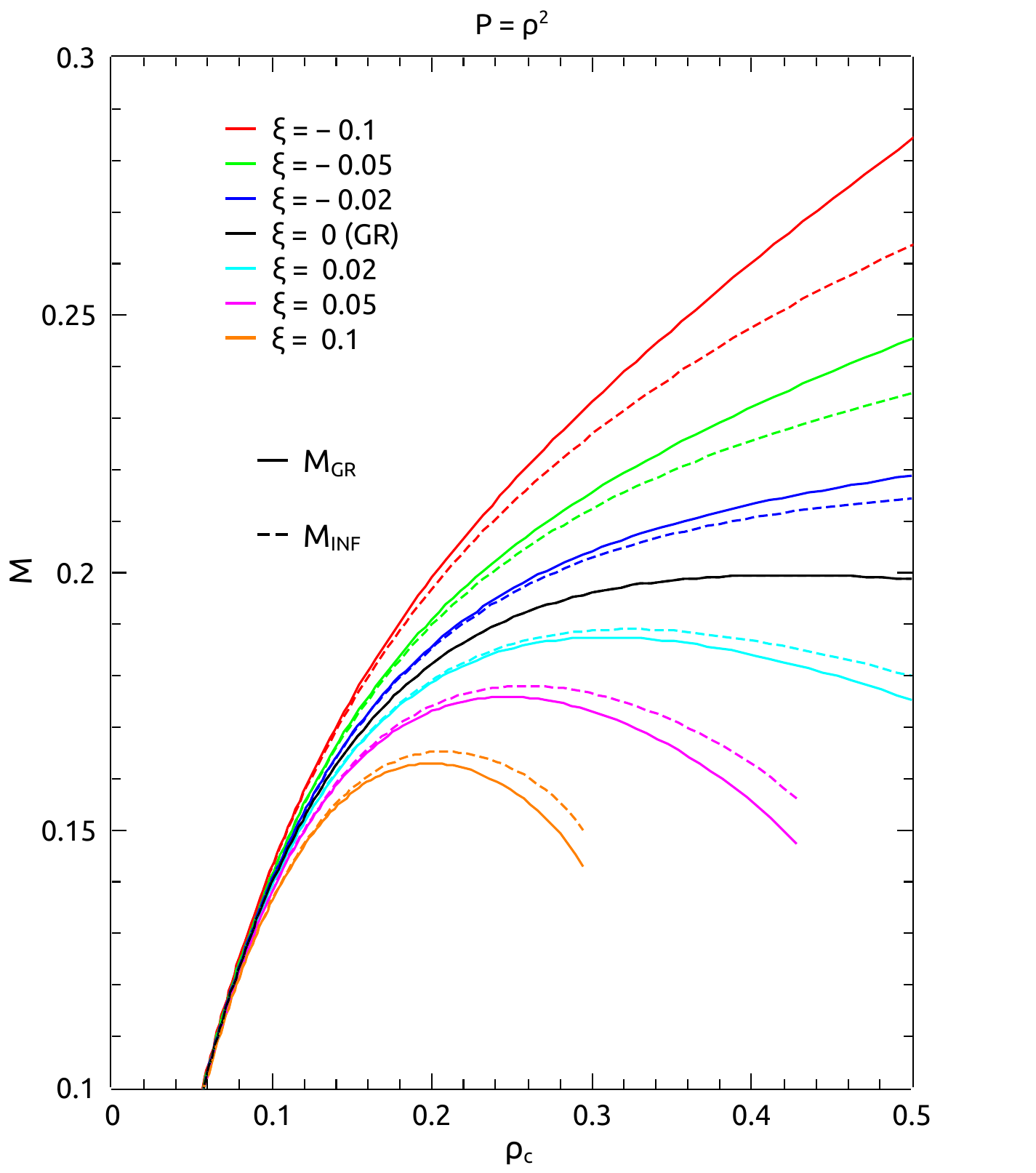}
    \includegraphics[scale=0.24]{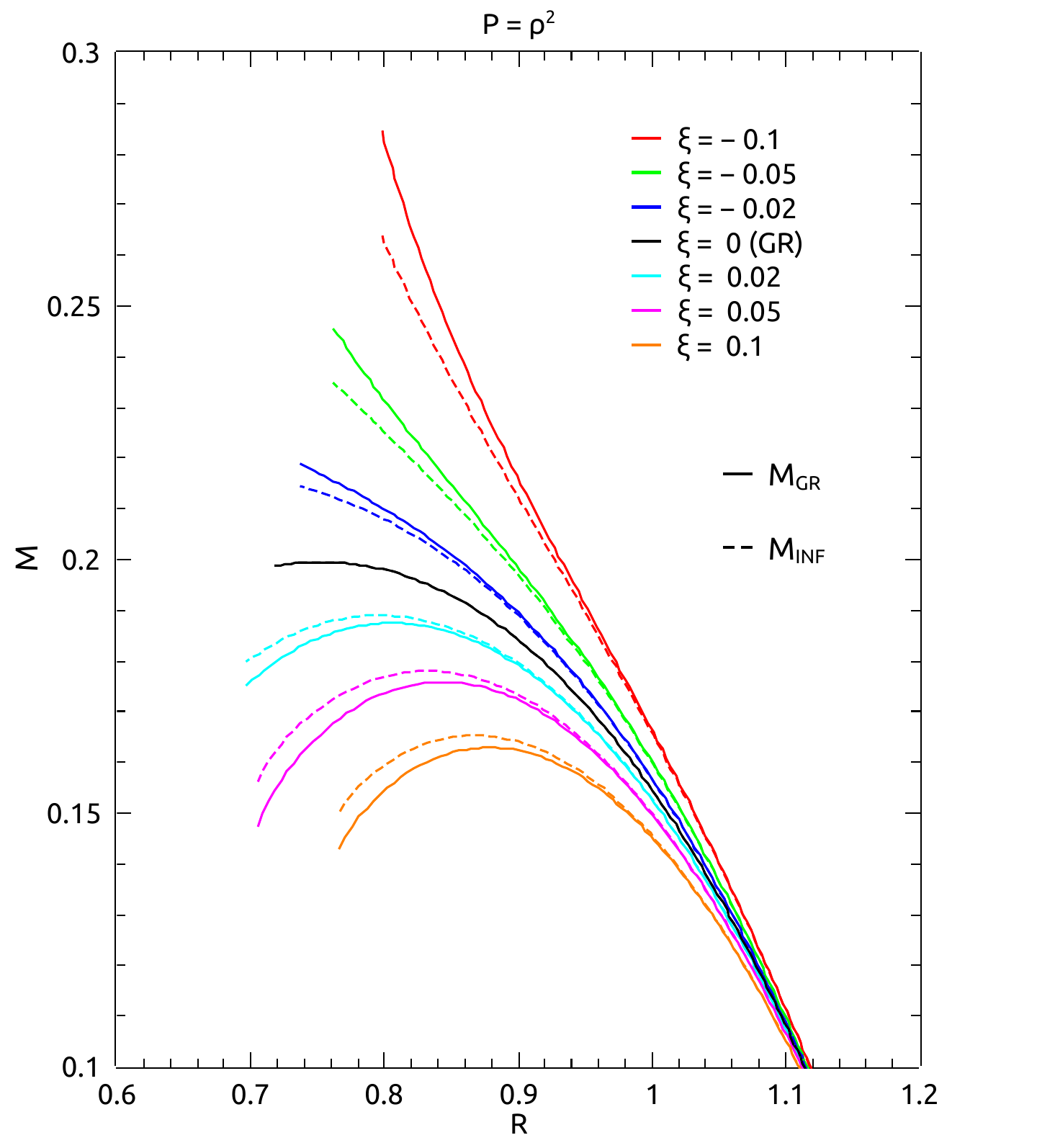}
    \caption{Left (right): Sequences of ${M_{GR}}$ {and} ${M_{I\! N\! F}}$ as a function of the central mass-energy density $\rho_c$ (radius R) for $P= \rho^2$ and different values of $\xi$.}
    \label{PPN1MD}
\end{figure*}

In Figure \ref{PPN1MD}, on the left panel, we plot sequences of ${M_{GR}}$ {and} ${M_{I\! N\! F}}$ as a function of the central mass-energy density $\rho_c$ for the $P= \rho^2$ EOS and different values of $\xi$.

Note that for $\xi \ge 0$ there are clearly maximum masses ${M_{GR}}$ {and} ${M_{I\! N\! F}}$, just like in GR. In addition, the greater $\xi$ is, the lower the maximum ${M_{GR}}$ {and} ${M_{I\! N\! F}}$. The calculations suggest that the maximum masses monotonically decrease for increasing values of $\xi$. One could think of the effective absolute value of the torsion (``gravity'') decreases for increasing values of $\xi$. It is worth noticing that the torsion is negative. In \cite{F-A1}, we show this explicitly. 

{Note that depending on the values of $\xi$ and $\rho_c$, ${M_{GR}}$ {and} ${M_{I\! N\! F}}$ differ significantly. We argue that ${M_{I\! N\! F}}$ would give the  more appropriate way to define mass because of its own definition, namely, the mass measure by an observer at infinity. This resembles the interpretations one gives for the mass that appears in Schwarzschild metric. Anyway, in the various calculations and figures that follow, we also keep the curves for $M_{GR}$ for the sake of comparisons.}

Considering now that $\xi$ could have negative values, one clearly sees that, as compared to GR, the masses for a given $\rho_c$ increases for decreasing values of $\xi$. The effective absolute value of the torsion (``gravity'') increases for decreasing values of $\xi$.

It is worth mentioning that we adopted a maximum value of $\rho_c$, which is the value for which the sound speed is equal to the speed of light, namely, $\rho_c^{max}= [n/(n+1)]^n$. For $P = \rho^2$, the maximum value of central density is $\rho_c = 1/2$.

It is evident, for the values of $\xi < 0$ presented in Figure \ref{PPN1MD}, that these sequences do not have turning points as those for $\xi \ge 0$. Consequently, there is no maximum mass for $\xi < 0$. Also, our calculations suggest that the masses increase monotonically for decreasing values of $\xi$.

{In \cite{Ilijic18}, problems of numerical instabilities are reported for certain combinations of parameters for both positive and negative $\xi$.} In our calculations, numerical instabilities also occur for certain combinations of parameters for $\xi >0$. On the other hand, for $\xi < 0$
our calculations do not have any problem of convergence.
This is probably because our system of equations is simpler than that of \cite{Ilijic18}, since we do not need to integrate a differential equation for $A$. As already mentioned, in our system of equations what appears is $A'$, which can be algebraically obtained by combining equations (\ref{press}) and (\ref{ce1}). {In conclusion, our novel approach is more robust than that of \cite{Ilijic18}, since it improves their numerical studies.}

Another way of presenting the sequences is to plot ${M_{GR}}$ {and} ${M_{I\! N\! F}}$ as a function of the radius $R$. These sequences, also for  $P= \rho^2$ EOS and different values of $\xi$, are shown in Figure \ref{PPN1MD}, on the right panel. Notice also that the compactness (i.e., $M/R$) increases for decreasing values of $\xi$. {We will come back to this issue of compactness later.}

It is also interesting to see on how a stiffer (softer) EOS as compared to $P = \rho^2$ is affected by different values of $\xi$. {This is an important issue that has not yet been addressed in the literature.}

\begin{figure*}
   \centering
    \includegraphics[scale=0.24]{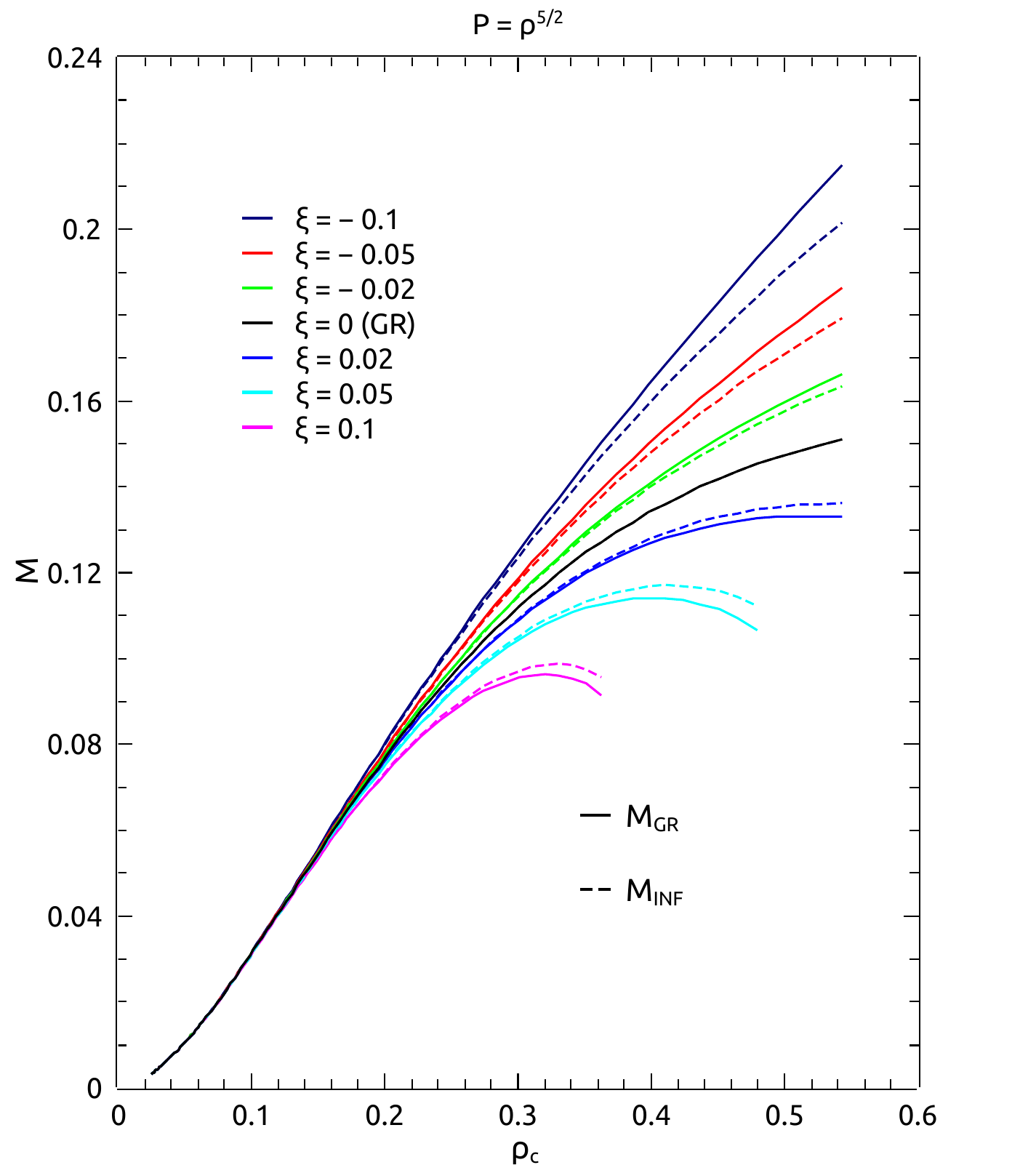}
    \includegraphics[scale=0.24]{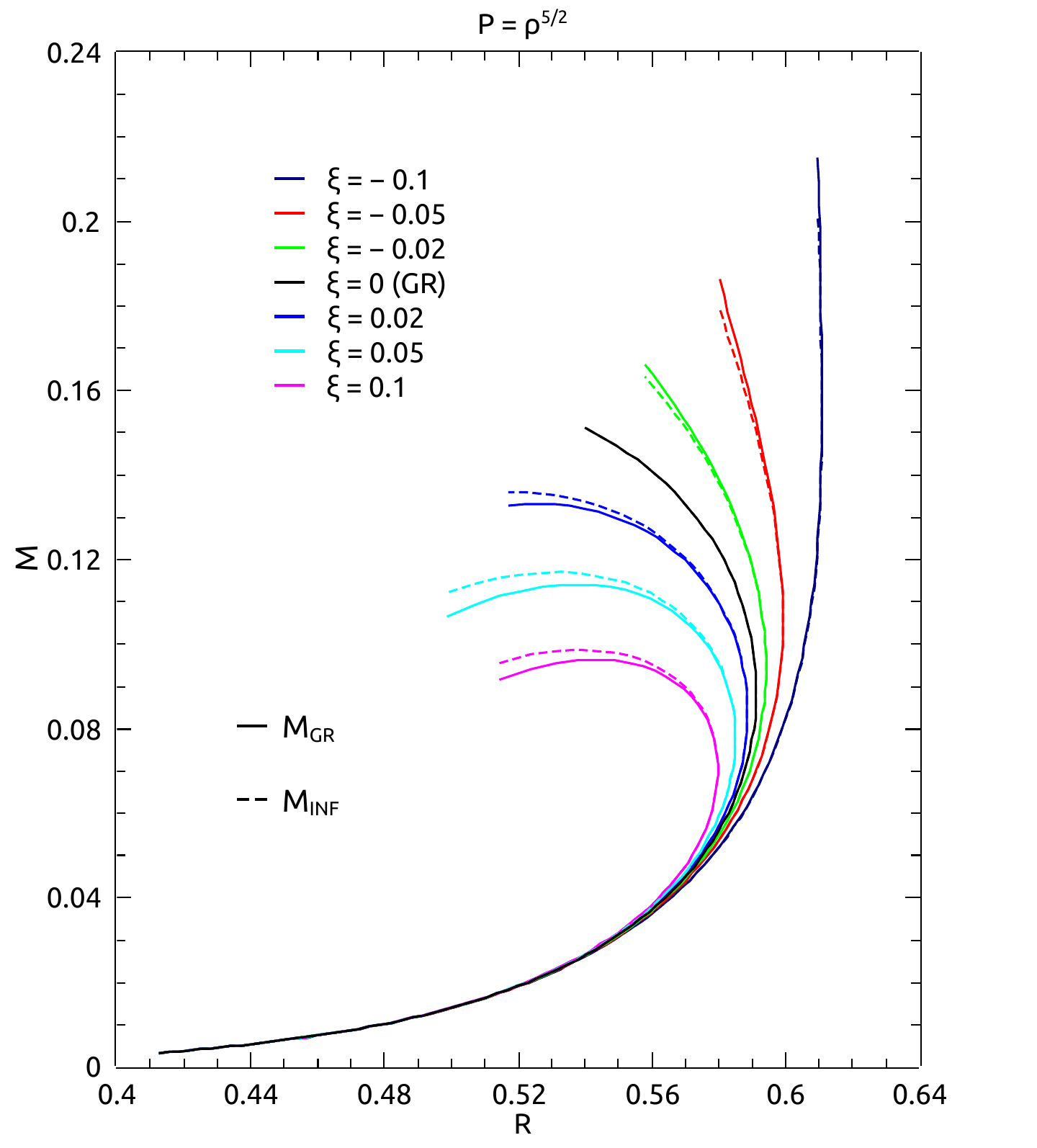}
    \caption{The same as Figure \ref{PPN1MD} now for $P= \rho^{5/2}$.}
    \label{PPN2s3MD}
\end{figure*}

In Figure \ref{PPN2s3MD}, it is plotted the same as in Figure \ref{PPN1MD} now for $n = 2/3$, that is, $P = \rho^{5/2}$. For $\xi >0$, the sequences for $n = 2/3$ have their maximum masses more significantly reduced as compared to GR than  sequences for $n=1$. 

On the other hand, for $\xi <0$, the sequences for $n = 2/3$ and $n=1$ have their maximum masses almost equally affected as compared to GR sequences. As for $n=1$, we adopted a maximum value for $\rho_c$, namely, $\rho_c^{max} = (2/5)^{2/3}$.  

{An interesting fact for the $P = \rho^{5/2}$ EOS is that ${M_{GR}}$ {and} ${M_{I\! N\! F}}$ 
do not differ significantly for the values of $\xi$ studied in this paper.}

\begin{figure*}
    \centering
    \includegraphics[scale=0.24]{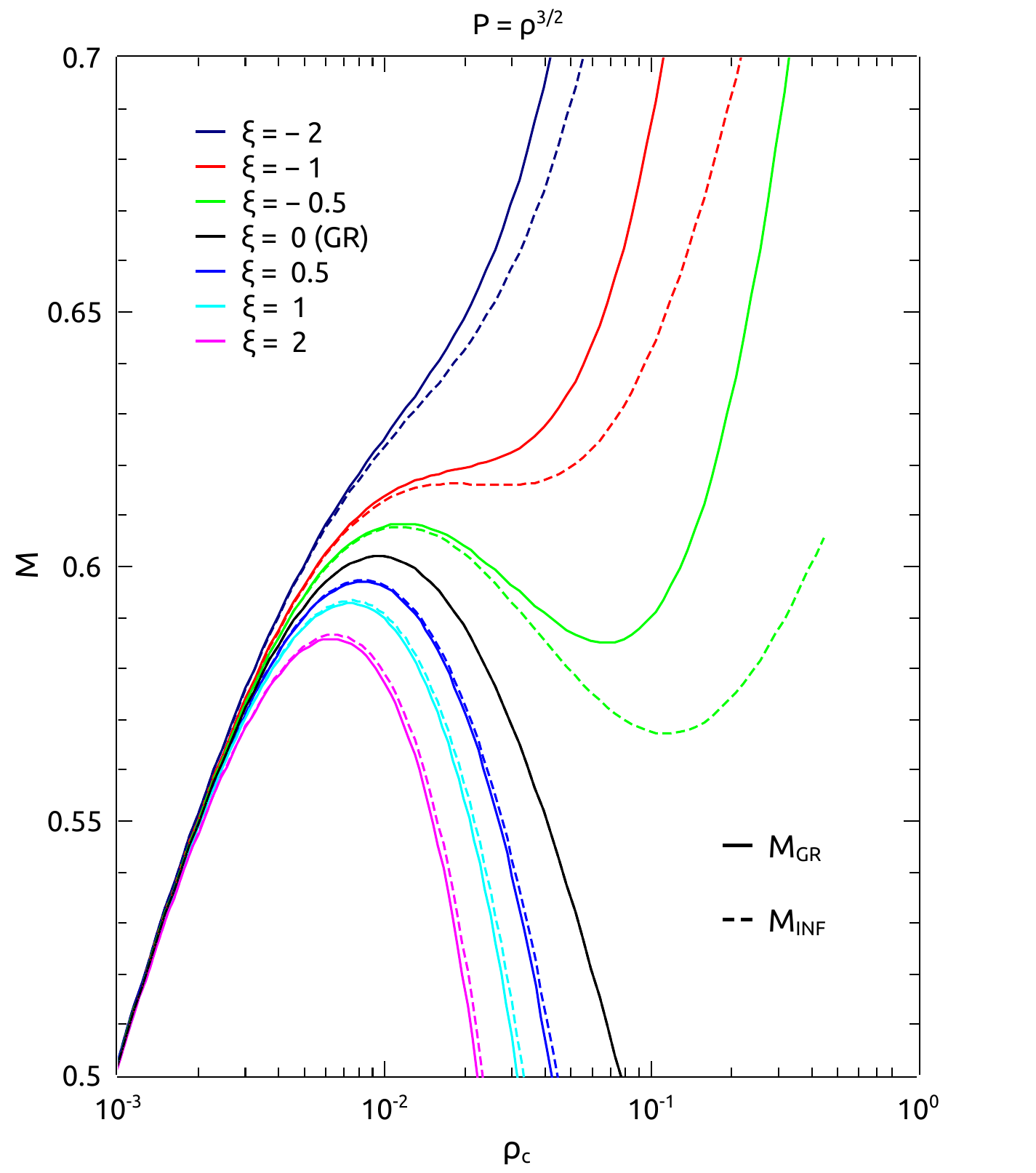}
    \includegraphics[scale=0.24]{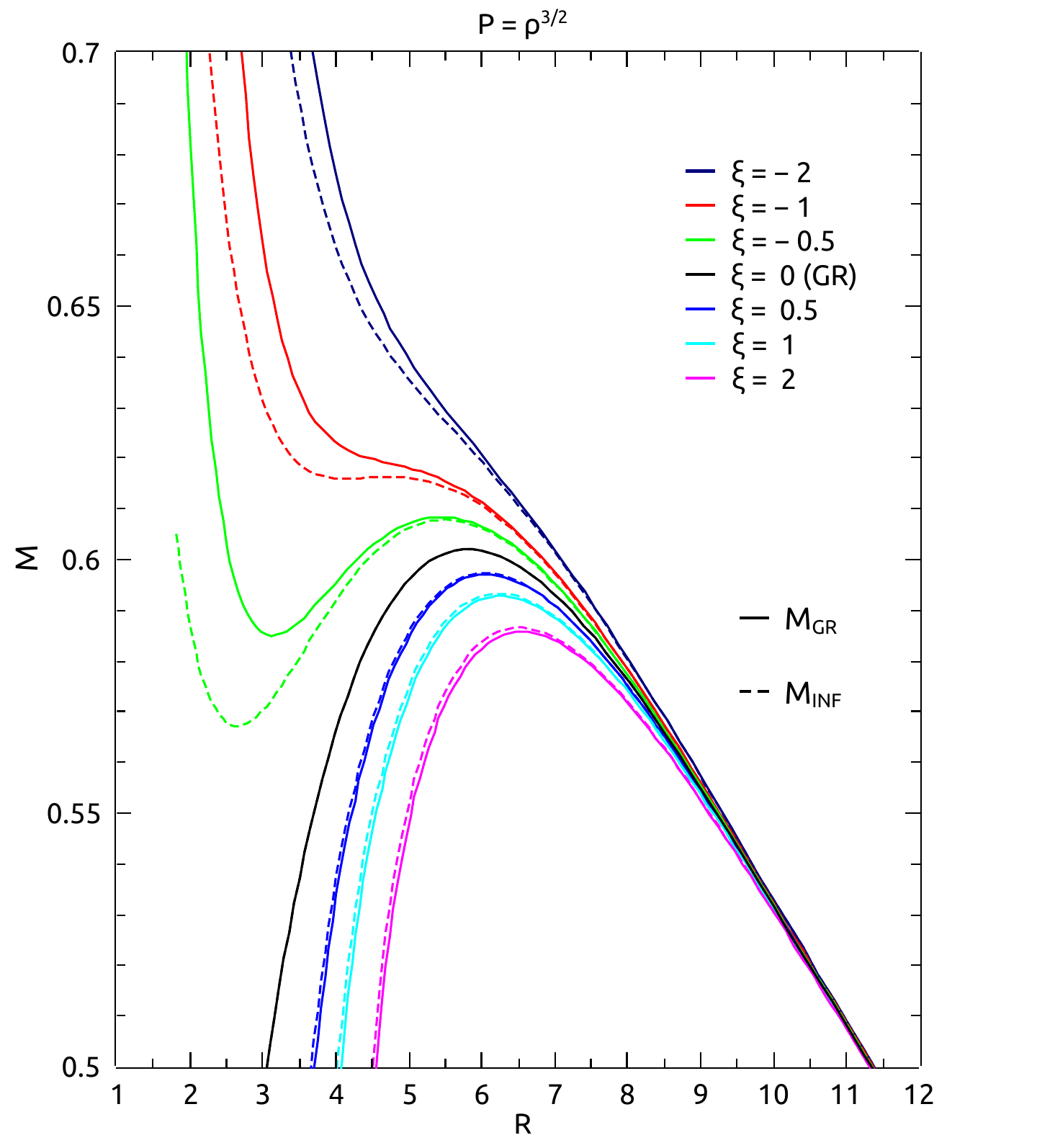}
    \caption{The same as Figure \ref{PPN1MD} now for $P= \rho^{3/2}$.}
    \label{PPN2MD}
\end{figure*}

An example of a polytropic EOS softer than $n=1$ can be obtained by choosing $n=2$, that is, $P = \rho^{3/2}$. The maximum value for $\rho_c$ is now $\rho_c^{max} = 4/9$.
In Figure \ref{PPN2MD}, it is plotted the same as in Figure \ref{PPN1MD} now for $n = 2$. 

Notice that to find significant differences for the GR sequence, higher values of $\mid\xi\mid$ are adopted. Thus, regarding polytropic EOSs, the softer ones are less sensitive to $\xi$ as compared to the stiffer ones. 

{The models for $P = \rho^{3/2}$ EOS has some interesting characteristics. Note that for $\xi > 0$, ${M_{GR}}$ {and} ${M_{I\! N\! F}}$ are not significantly different. The model for $\xi = - 0.1$ has a local maximum around $\rho_c \sim 0.01$ and a local minimum around $\rho_c \sim 0.1$. An interesting question to ask is whether or not stars modelled for densities $\rho_c \gtrsim 0.01$ are unstable. The issue related to stability of stars in $f(T)$ deserves to be considered in the literature, but it is out of the scope of this article.}

\begin{figure*}
    \centering
    \includegraphics[scale=0.24]{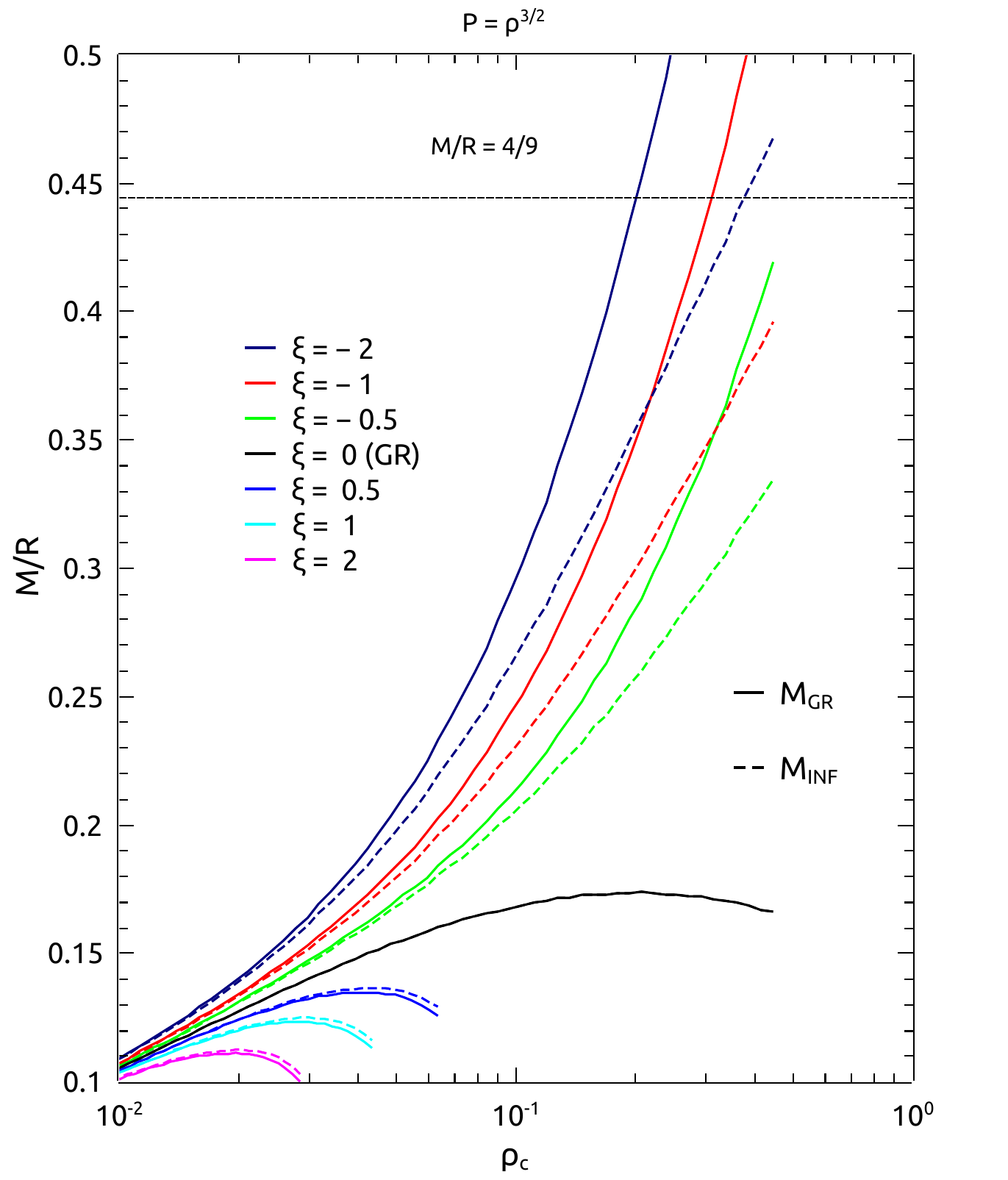}
    \includegraphics[scale=0.24]{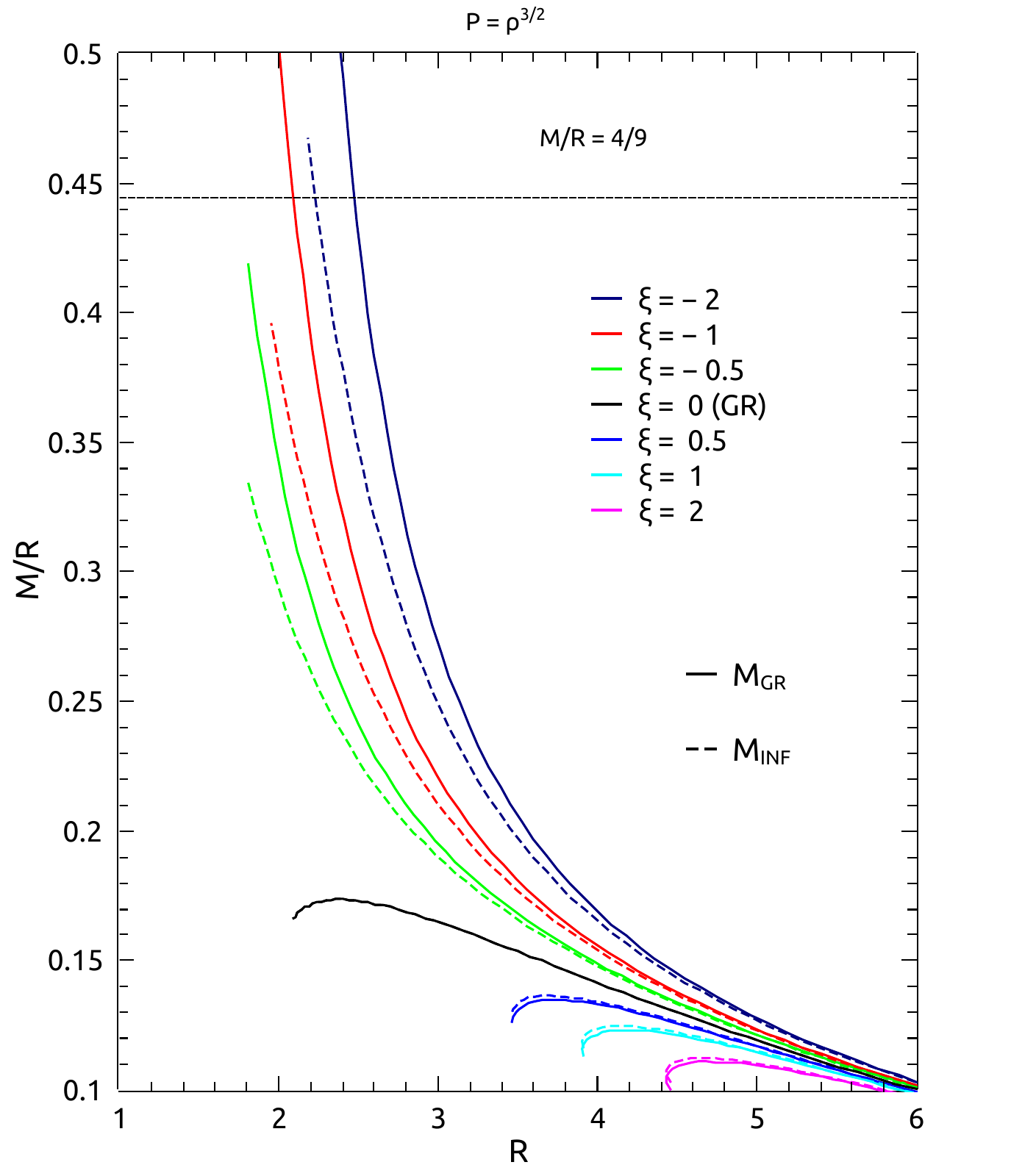}
   \caption{Left (right): Sequences of M/R as a function of $\rho_c$ (R) for $P= \rho^{3/2}$ and different values of $\xi$.}
    \label{PPN2MRD}
\end{figure*}
{As already mentioned above, negative/positive values of $\xi$ lead to more/less compact stars than those of the RG ones. This can be seen in figures \ref{PPN1MD} - \ref{PPN2MD}. However, the best way to see the behaviour of compactness for different values of $\xi$ is to plot sequences of M/R (compactness) as a function of $\rho_c$ and/or R. {It is worth noting, to the best of our knowledge, that there are no studies in the literature that consider the compactness of stars in $f(T)$ gravity.} 

As a representative example, we consider the polytropic EOS  $P= \rho^{3/2}$. It is worth mentioning that the other polytropics considered here have a similar behaviour.}

{In Figure \ref{PPN2MRD}, we show the compactness sequences ``M/R $\times$ $\rho_c$" and ``M/R $\times$ R" for $P= \rho^{3/2}$ and different values of $\xi$.
{Note that the compactness is presented for both ${M_{GR}}$ {and} ${M_{I\! N\! F}}$.}
For reference, it is indicated the M/R = 4/9 line, which is the maximum compactness of a star in GR (Buchdahl limit). Notice that the plots are limited to M/R = 1/2, which in GR is the compactness of a black hole.}

{As a general conclusion, note that the more negative $\xi$ is, the greater the compactness. Depending on the value of $\xi$, the 4/9 (Buchdahl limit) and 1/2 (black hole) compactness of GR can be exceeded.}

{A relevant question that arises is what is the limit of the compactness of stars in the $f(T)$ gravity, in particular in the one studied here. A relevant issue is if stars with compactness $M/R > 4/9$ can really exist in $f(T)$.} {This is for sure an issue that must be properly addressed and is closely related to the stellar stability, but it is not addressed here since is out of the scope of this paper.}

\section{Final Remarks}\label{sec 4}

{In our previous article, we have shown that it is possible to obtain equations to model compact stars in $f(T)$ gravity using a more general approach than others found in the literature.}

{The key point of our approach is to realise that the set of equations (\ref{press}), (\ref{BL}), (\ref{ce1}) and (\ref{dmdr}) is the one to be numerically solved.
It is noteworthy that the combination of equations (\ref{press}) and (\ref{ce1}) works like a generalisation of the TOV GR differential equation for $P(r)$. Notice also that in GR the function $B(r)$ is closely related to $m(r)$, namely, $e^{-B} = 1 - m(r)/r$. Here, to obtain $B(r)$, one has to numerically integrate equation (\ref{BL}). Although this equation is complicated, it is numerically treatable.}

{Our novel approach can be considered an improvement of that of the paper \cite{Ilijic18}, since we can model stars for any central density for $\xi < 0$ without any the problems of convergence they faced. This is probably because our system of equations is much simpler and numerical stable. In their system of equations there is a differential equation for $A'$, which is very complicated. In our approach, given that we use the {\it conservation equation}, it is not necessary to use a differential for $A'$, since this function can be obtained algebraically.} 

{One could ask why study polytropic EOSs. First, since \cite{F-A1} and \cite{Ilijic18} consider together only two distinct polytropic indices, there is room in the literature to explore other ones. Second, polytropic EOSs of different index could be useful in the study of stellar stability. The stability window may well depend on the polytropic index. }

{Notice that we do not consider in the present paper the stability of the solutions. This is an important issue that should be addressed but is out of our objectives here.}

{In addition, we have considered the analysis also using the mass measured by an ``observer at infinity'', which seems to be the most appropriate mass definition in this case, rather than that from GR.}

A relevant issue has to do with the maximum compactness of stars in $f(T)$ gravity, which is closely related to stellar stability too.
Our calculations show that the compactness can be higher than 4/9 (maximum compactness of a star in GR), and it can even be bigger than the compactness of a black hole, which is of 1/2. Notice that there is not a similar study in the literature via sequences of models such as presented in Figure \ref{PPN2MRD}. 

There is certainly a limit to compactness, which studies related to stellar stability in $f(T)$ gravity should also be able to shed light.

The obvious next step is to consider realistic EOSs, {which unlike politropic EOSs, take in to account the nuclear matter inside the neutron star.} This issue was recently considered by us in \cite{F-A2}.

{Another important issue to be stressed is that, although we consider in the present paper that $f(T) = T + \xi T^2$, one can in principle consider any $f(T)$ function. We will also consider this issue in the near future.}

\bmhead{Acknowledgments}
J.C.N.A. thanks  CNPq (308367/2019-7) for partial financial support. The authors would like to thank Rafael da Costa Nunes for discussions related to the f(T) theory. {Last but not least, we thank the referee for the valuable review of our article, which helped to improve it.}

\section*{Declarations}

\par\noindent{\bf Conflict of interest/Competing interests} The authors declare no conflict of interest/competing interests.

\par\noindent{\bf Data Availability Statement} No Data associated in the manuscript.

\end{document}